\documentclass[showpacs,preprintnumbers,amsmath,amssymb,twocolumn]{revtex4}
\usepackage{bm}
\usepackage{graphicx}
\usepackage{subfigure}

\begin{document}

\title{Lagrangian studies in convective turbulence}
\author{J\"org Schumacher\footnote{email: joerg.schumacher@tu-ilmenau.de}}
\affiliation{Institut f\"ur Thermo- und Fluiddynamik,
                   Technische Universit\"at Ilmenau, Postfach 100565, D-98684 Ilmenau, Germany}
\date{\today}

\begin{abstract}
We present high-resolution direct numerical simulations of turbulent three-dimensional 
Rayleigh-B\'{e}nard convection with a focus on the Lagrangian properties of the flow. 
The volume is a Cartesian slab with an aspect ratio of four bounded 
by free-slip planes at the top and bottom and with periodic side walls. The turbulence 
is inhomogeneous with respect to the vertical direction. This manifests in different 
lateral and vertical two-particle dispersion and in a dependence of the dispersion
on the initial tracer position for short
and intermediate times. Similar to homogeneous isotropic turbulence, the dispersion 
properties depend in addition on the initial pair separation and yield a short-range Richardson-like 
scaling regime of two-particle dispersion for initial 
separations close to the Kolmogorov dissipation length. The Richardson constant is about
half the value of homogeneous isotropic turbulence. The multiparticle statistics is very close  to the 
homogeneous isotropic case. Clusters of four Lagrangian tracers show a clear trend to form flat, 
almost coplanar objects in the long-time limit and deviate from the Gaussian prediction. 
Significant efforts have been taken to resolve the 
statistics of the acceleration components up to order four correctly. We find that the
vertical acceleration is less intermittent than the lateral one.  The joint statistics 
of the vertical acceleration with the local convective  and conductive heat flux suggests that rising
and falling thermal plumes are not associated with the largest acceleration magnitudes. It turns out also
that the Nusselt number which is calculated in the Lagrangian frame converges slowly in time to the 
standard Eulerian one.      
\noindent 
\end{abstract}

\maketitle

\section{Introduction}
Turbulent convection is one of the best studied fundamental flows in fluid dynamics research
\cite{Kadanoff2001,Ahlers2009}. One reason is the large range of examples and applications  
in nature and technology for which a turbulent motion is initiated and sustained by heating a fluid 
from below and cooling from above. Almost all of these studies have been conducted 
in the Eulerian frame of reference. They were primarily focussed to the mechanisms  of local 
\cite{Shishkina2007,Zhou2007,Shishkina2008,Emran2008} and global 
\cite{Kerr1996,Niemela2000,Amati2005,Funfschilling2005} turbulent heat transfer.

The Lagrangian perspective of turbulence, in which the fields are monitored along the trajectories of
infinitesimal fluid parcels, has recently produced new insights into
the local topology of fluid parcel tracks, the local strength of accelerations,  and the statistics
of time increments of turbulent fields \cite{Toschi2009,Schumacher2009}. The progress is caused on the one hand 
by significant innovations in the experimental techniques, such as three-dimensional particle tracking 
\cite{Mann2000,LaPorta2001,Guala2005} or acoustic methods \cite{Mordant2001}.  On the other
hand, direct numerical simulations of turbulence become now feasible that resolve three-dimensional
Lagrangian turbulence at moderate and higher Reynolds numbers 
\cite{Yeung2002,Boffetta2002,Biferale2005,Sawford2008}.  Both, experiments and simulations,  made 
a deeper understanding of the small-scale intermittency and its connection with large accelerations 
of fluid parcels possible. 

Lagrangian investigations
in convective turbulence are however rare. Several reasons can be given for this circumstance. 
First, on the experimental side it is desirable to monitor the temperature along the particle tracks 
beside the velocity components and the accelerations. Only recently, Gasteuil {\it et al.} 
\cite{Gasteuil2007} constructed therefore a smart particle, that monitors velocity, temperature and 
orientation while moving through the cell. Due to the integrated power supply the particle diameter remained 
however larger than the thermal boundary layer thickness, such that the large-scale bulk motion can be 
monitored only. Second, it is also clear that the complexity of direct numerical simulations increases
since the temperature field has to be advected in addition to the velocity. Temperature tracking
along the tracer positions requires additional interpolations. Furthermore, one cannot return to simulations in a
fully periodic cube, the so-called homogeneous Rayleigh-B\'{e}nard convection setup, since
the periodicity in the direction of the mean temperature gradient causes a self-amplifying fluid
motion. This was discussed in detail by Calzavarini {\it et al.} \cite{Calzavarini2005,Calzavarini2006}.  
Third, the turbulence is inhomogeneous --at least in the vertical direction as in the following
setup-- and it is thus not clear which of
findings from the homogeneous isotropic purely hydrodynamic turbulence pertain. For
example, the height dependence of the statistics has to be considered additionally.

First numerical attempts have been made recently to study some aspects of the heat transfer and 
tracer dispersion in the Lagrangian framework of convective turbulence \cite{Schumacher2008}. 
The motivation of the study can be condensed in one question: Which new insight into the nature
of turbulent convection provides the complementary Lagrangian view?  One result of \cite{Schumacher2008} 
was to determine a mixing zone which is dominated by rising and falling thermal plumes. This is done by 
combining acceleration and local convective heat flux statistics. The mixing zone starts right
above the thermal boundary layer and extends several tens of the boundary layer thickness 
into the bulk of the cell. 
Thermal plumes are fragments of the thermal boundary layer that detach in the
vicinity of the top and bottom isothermal planes. The existence of a mixing zone has 
been suggested in several Eulerian studies on the basis of other criteria, 
e.g. \cite{Castaing1989,Xia2002} and was thus confirmed in the complementary Lagrangian 
frame of reference \cite{Schumacher2008}.  

The present work extends the previous study \cite{Schumacher2008} into several directions. 
Beside the local convective,  the local conductive heat flux is studied along the tracer tracks. 
It requires to monitor temperature gradient components. Furthermore, the analysis of the Lagrangian tracer
dispersion is extended. In addition to the hydrodynamic case \cite{Sawford2008}, we study the dependence 
of pair dispersion on the initial separation {\em and} the initial seeding position. 
As discussed in Refs. \cite{Chertkov1999,Pumir2000,Biferale2005} for the pure hydrodynamic case, higher order particle 
statistics requires to track little clusters of tracers. We provide here an analysis of 
the four-particle-statistics, where
the tracers start out of groups of tetrahedra of different sidelengths and initial vertical positions.   

The outline of the manuscript is as follows. In the next section the equations of motion,  the numerical scheme, 
the Lagrangian tracer tracking and the turbulent heat transfer. In section III, some results of the Eulerian 
statistics of the temperature field are presented. This section is followed by sections on the Lagrangian particle 
dispersion,  the acceleration statistics and the conductive and convective heat flux. We 
conclude with a short discussion of our results and will give a brief outlook to possible extensions of the 
present work.

\section{Numerical model}
\subsection{Equations of motion and boundary conditions}
The Boussinesq equations, i.e. the Navier-Stokes equations for an incompressible flow
with an additional buoyancy term  $\alpha g \theta {\bm e}_z$ and the advection-diffusion 
equation for the tempe\-rature field, are solved by a standard pseudospectral method for 
the three-dimensional case \cite{Schumacher2007}. 
The equations are given by
\begin{eqnarray}
{\bm \nabla\cdot \bm u}&=&0\,,\\
\frac{\partial {\bm u}}{\partial t}+({\bm u\cdot \bm\nabla}){\bm u}&=&
-{\bm\nabla} p +\nu {\bm\nabla}^2{\bm u}+\alpha g \theta {\bm e}_z\,,\\
\frac{\partial \theta}{\partial t}+({\bm u\cdot\bm \nabla})\theta &=&
\kappa {\bm\nabla}^2\theta+u_z \frac{\Delta T}{H}\,.
\end{eqnarray}
Here, ${\bm u}$ is the turbulent velocity field, $p$ the (kinematic) pressure field and $\theta$ the temperature 
fluctuation field. The system parameters are: gravity acceleration $g$, kinematic viscosity 
$\nu$, thermal diffusivity $\kappa$, vertical temperature gradient 
$\Delta T/H$, and thermal expansion coefficient $\alpha$. The temperature field is 
decomposed into a linear profile and fluctuations $\theta$ about the profile
\begin{equation}
T({\bm x},t)=-\frac{\Delta T}{H}(z-H/2)+\theta({\bm x},t)\,.
\label{reynolds0}
\end{equation}
Since $T$ is prescribed and constant at bottom and top boundaries $z=0$ and $z=H$, the condition $\theta=0$ follows there. 
Here, $\Delta T >0$. The dimensionless control para\-meters are the Prandtl number $Pr$, the Rayleigh number 
$Ra$, and the aspect ratio $\Gamma$,
\begin{eqnarray}
Pr&=&\frac{\nu}{\kappa}\,,\\
Ra&=&\frac{\alpha g H^3 \Delta T}{\nu\kappa}\,,\\
\Gamma&=&\frac{L}{H}\,.
\end{eqnarray}
The simulation domain is $V=L\times L\times H=[0,\Gamma\pi]\times[0,\Gamma\pi]\times[0,\pi]$.
In lateral directions $x$ and $y$, periodic boundary conditions are taken.
In the vertical direction $z$, free-slip boundary conditions are used which are given by
\begin{eqnarray}
u_z=\theta=0 \;\;\;\;\mbox{and}\;\;\;\;\partial_z u_x=\partial_z u_y=0\,.  
\end{eqnarray}
The computational grid has a size of $N_x\times N_y\times N_z=2048\times 2048\times 513$ points. For an aspect 
ratio $\Gamma=4$, it is thus equidistant in all three space directions with a grid spacing $\Delta x$. 
Time-stepping is done 
by a second-order predictor-corrector scheme.  The production runs are conducted on one rack 
of the Blue Gene/P system which corresponds with 4096 MPI tasks \cite{Schumacher2007}. 
We use volumetric Fast Fourier  Transforms based on the p3dfft package by D. Pekurovsky 
\cite{Pekurovsky2008}. The spectral resolution is $k_{max}\eta_K=4.5$ where $k_{max}=2\sqrt{2}\pi 
N_x/(3 L_x)$.  Quantity $\eta_K=\nu^{3/4}/\langle\epsilon\rangle^{1/4}$ is the Kolmogorov scale with the
mean energy dissipation rate $\langle\epsilon\rangle$.

\begin{figure}
\centerline{\includegraphics[scale=0.5]{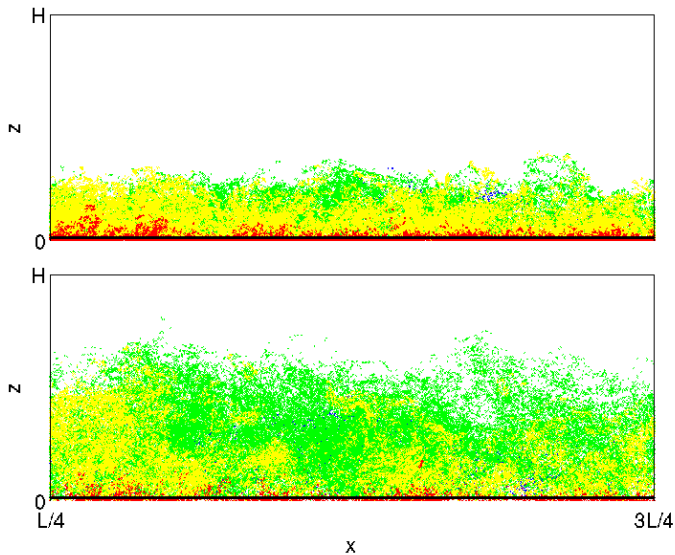}}
\caption{(Color online) Side view of two instantaneous Lagrangian tracer distributions at $t=9.5\tau_{\eta}$ (top)
and $t=19\tau_{\eta}$ (bottom). The tracers are colored with respect to the local temperature $T$ at their position.
Four intervals are taken: $T\in[\Delta T/2,\Delta T/3]$, $T\in[\Delta T/3,\Delta T/6]$, $T\in[\Delta T/6,0]$ and for 
$T\in[0,-\Delta T/6]$. The whole ensemble consists of $2\times 10^5$ tracers.  Tracers started in the
whole $x-y$ plane at the height of the thermal boundary layer thickness which is indicated by the solid
line very close to the bottom plane in both plots.}
\label{fig0}
\end{figure}

\subsection{Lagrangian particle tracking}
Lagrangian tracer particles follow the streamlines of the turbulent velocity field 
in correspondance with
\begin{equation}
\dot{\bm x}={\bm u}({\bm x}(t),t)\,.
\end{equation}
For the majority of the analysis, we seeded $3\times 10^5$ tetrahedra aligned along the outer 
coordinate axes in the box, i.e. ${\bm x}_1={\bm x}_0$, ${\bm x}_2={\bm x}_0+{\ell} {\bm e}_x$, 
${\bm x}_3={\bm x}_0+{\ell} {\bm e}_y$, and ${\bm x}_4={\bm x}_0+{\ell} {\bm e}_z$. The vector 
${\bm x}_0$ is randomly chosen in the box. The tracer ensemble was divided into 6 six groups 
with initial sidelengths of ${\ell} =1, 2, 4, 8, 16$ and  32  grid spacings $\Delta x$ which correspond 
with 0.5, 1, 2, 4, 8 and 16 $\eta_K$. The two-particle dispersion analysis is consequently conducted 
for the three tracer pairs $\{{\bm x}_1,{\bm x}_2\}$, $\{{\bm x}_1,{\bm x}_3\}$, and $\{{\bm x}_1,{\bm x}_4\}$
of each tetrahedron.

For the two- and multiparticle statistics, we run in addition a 
simulation with the following initial conditions:  again  ${\bm x}_1={\bm x}_0$, ${\bm x}_2={\bm x}_0+
{\ell} {\bm e}_x$, ${\bm x}_3={\bm x}_0+{\ell} {\bm e}_y$, and ${\bm x}_4={\bm x}_0+
{\ell} {\bm e}_z$. The $x$ and $y$ coordinates of the vector ${\bm x}_0$ are again randomly 
chosen. The vertical coordinate corresponds with $z_0=\delta_T/2$, $\delta_T$,  $10\delta_T$, 
$20\delta_T$ and $H/2$. Here, we pick ${\ell}=\eta_K/2$ and $2\eta_K$. 

The Lagrangian particles are advanced in time simultaneously with the Boussinesq equations. 
The velocity, temperature and temperature gradient components  at intergrid positions are calculated 
by trilinear interpolation. The full particle set is written out each 0.45 $\tau_{\eta}$. Here, $\tau_{\eta}=
\sqrt{\nu/\langle\epsilon\rangle}$ is the Kolmogorov time. Accelerations along the 
Lagrangian tracks are calculated from three successive 
integration steps ($\Delta t=0.006\tau_{\eta}$) and the output interval is the same as for the particle 
positions, velocities, temperature, and temperature gradient.  We thus gather Lagrangian statistics 
over up to $4.8\times 10^8$ tracer particle events. Figure \ref{fig0} illustrates the initial phase of
the tracer dispersion. All tracers start from a $x-y$ plane close to the bottom wall.

\subsection{Turbulent heat transfer}
The convective turbulence is studied for one parameter setting. The Rayleigh number is $Ra=1.2\times 10^8$, the 
Prandtl number $Pr=0.7$ and the aspect ratio $\Gamma=4$. The response of the system is a turbulent heat transport
as quantified by the dimensionless (Eulerian) Nusselt number which is given for a plane at fixed height $z$ by
\begin{equation}
Nu(z)=\frac{\langle u_z T\rangle_{A,t}-\kappa \partial_z\langle T\rangle_{A,t}}{\kappa\Delta T/H}\,,
\label{Nulocal}
\end{equation} 
where $\langle\cdot\rangle_{A,t}$ denote averages in planes at $z$ and with respect to time.
The value of $Nu(z)$ is constant and independent of $z$. The global Nusselt number is then defined as 
\begin{equation}
Nu=\frac{1}{H}\int_0^H Nu(z) \mbox{d}z =1+\frac{H}{\kappa\Delta T}\langle u_z T\rangle_{V,t}\,,
\end{equation} 
\begin{figure}
\centerline{\includegraphics[scale=0.35]{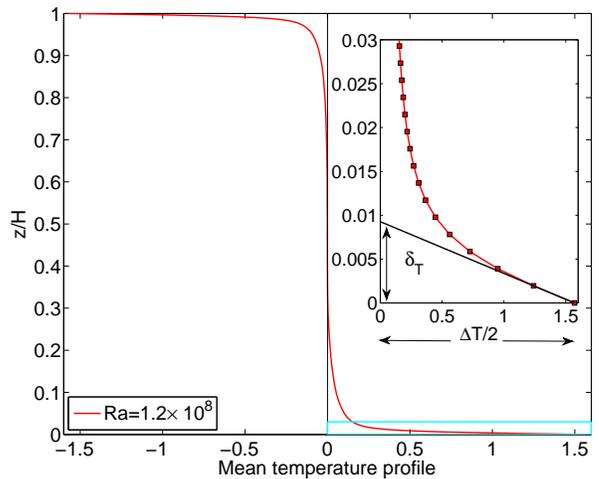}}
\caption{(Color online) Mean temperature profile $\langle T(z)\rangle_{A,t}$
of the turbulent convection run at $Ra=1.2\times 
10^8$ and $Pr=0.7$. The inset shows the resolution of the thermal boundary layer with 7 
grid planes. It also indicates the geometric interpretation of the thermal boundary layer 
thickness $\delta_T$ (see the text). The lower right box in the main figure indicates the size of the 
magnification.}
\label{fig1}
\end{figure}
\begin{figure}
\centerline{\includegraphics[scale=0.45]{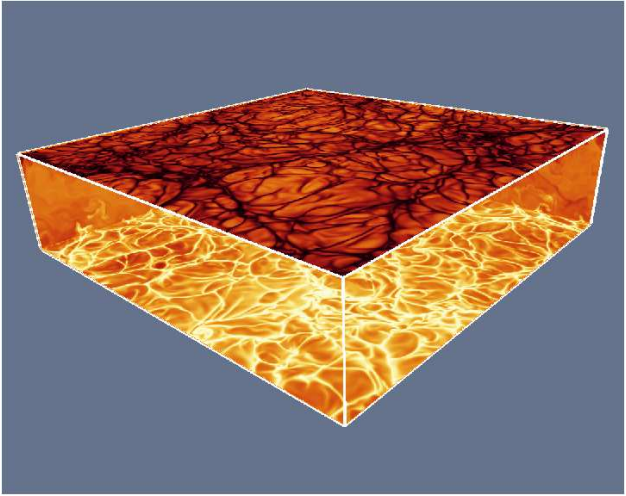}}
\caption{(Color online) Contour plot of the instantaneous temperature field $T({\bm x},t_0)$.}
\label{fig2}
\end{figure}
\begin{figure}
\centerline{\includegraphics[scale=0.45]{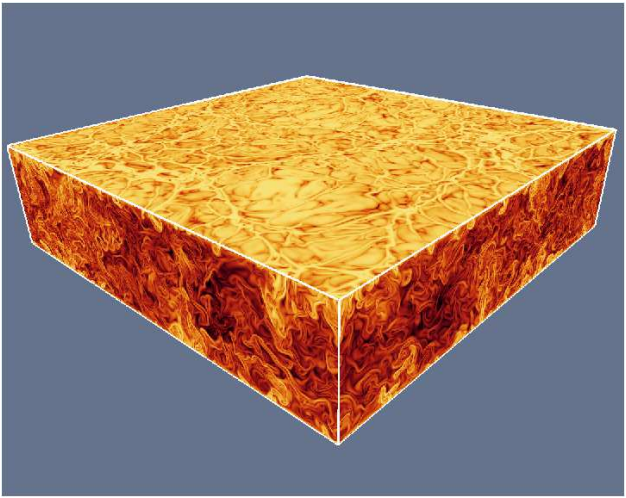}}
\caption{(Color online) Contour plot of the instantaneous thermal dissipation 
rate field $\chi({\bm x},t_0)$. Data correspond to those in Fig. \ref{fig2}. In order to highlight 
the small-amplitude dissipation filaments in the bulk, we plot contours of the 
decadic logarithm of $\chi$.}
\label{fig3}
\end{figure}
where $\langle\cdot\rangle_{V,t}$ is a combined volume and time average.
The Nusselt number for the present free-slip boundary case follows to $Nu=56.36\pm 0.59$. Similar to Julien {\it et al.} 
\cite{Julien1996}, we find an enhanced turbulent heat transport in comparison to no-slip top and bottom plates. For Rayleigh numbers
between $9.8\times 10^5$ and $1.2\times 10^8$, we fit the power law  $Nu=0.166\times Ra^{0.316}$ to the data.   

\section{Eulerian temperature statistics}
Figure \ref{fig1} displays the mean temperature profile as a function of height. The total temperature can take values
between $-\Delta T/2$ and $\Delta T/2$ only. As typical for higher Rayleigh numbers,  the jump of mean profile to zero is observed 
across a thin layer, the thermal boundary layer. The inset magnifies the vicinity of the bottom 
plate.  The thickness of the thermal boundary layer is defined as
\begin{equation}
\delta_T=\frac{H}{2 Nu}\,.
\end{equation} 
For $z=0$, the conductive part of (\ref{Nulocal}) contributes to $Nu$ 
only and we can set $Nu=-\kappa \partial_z\langle T\rangle_{A,t}|_{z=0}
/(\kappa\Delta T/H)$. This leads to $\delta_T=-\Delta T/(2\partial_z\langle T\rangle_{A,t}|_{z=0})$ and to the geometric
derivation of the thermal boundary layer thickness (as indicated in the inset of Fig. \ref{fig1}). In contrast to the no-slip case, we have 
$\langle u_x\rangle_{A,t}=\langle u_y\rangle_{A,t}=\langle u_z\rangle_{A,t}=0$ . Consequently, no velocity boundary 
layer is present.  The Taylor microscale Reynolds number $R_{\lambda}=\sqrt{5/(3\nu\langle\epsilon\rangle)}\,\langle 
u_i^2\rangle\approx 143$. 
\begin{figure}
\centerline{\includegraphics[scale=0.4]{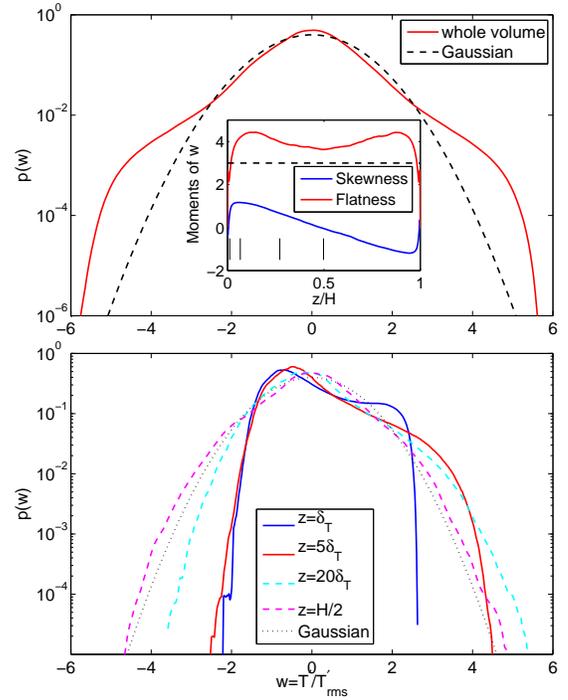}}
\caption{(Color online) Statistics of the temperature fluctuations $w=T^{\prime}/T^{\prime}_{rms}$. 
Top: Probability density function (PDF) of the temperature fluctuations. Data are compared with a 
Gaussian distribution (dashed line). Inset:  Skewness and flatness of the temperature fluctuations 
$w$ as a function of the vertical coordinate $z/H$. The profiles are obtained 
by averaging over lateral planes and a sequence of statistically independent snapshots. The Gaussian 
value for the flatness $F=3$ is indicated by the dashed line. Bottom: PDFs of the temperature fluctuations 
taken in different planes which are indicated in the legend and by vertical solid lines in the inset of the top figure.}
\label{fig4}
\end{figure}

Figure \ref{fig2} shows an instantaneous snapshot of the total temperature field $T({\bm x},t)$.  Contour 
plots in two sideplanes and close to the top and bottom planes are shown. We observe a typical feature
of thermal convection -- the ridge-like maxima which correspond with thermal plumes that detach randomly. They
form a skeleton which is advected by the flow close to the boundaries. The plumes coincide with local 
maxima of the thermal dissipation rate field (see Fig. \ref{fig3} and compare it with Fig. \ref{fig2}) which is 
defined as 
\begin{equation}
\chi({\bm x},t)=\kappa ({\bm\nabla}T^{\prime}({\bm x},t))^2\,.
\end{equation}
The definition contains the temperature fluctuations  which are given by
\begin{equation}
T^{\prime}({\bm x},t)=T({\bm x},t)-\langle T(z)\rangle_{A,t}\,.
\end{equation}
The probability density function (PDF) of $T^{\prime}$ is shown in Fig. \ref{fig4}. We compare the PDF 
of data taken from the whole slab volume with the Gaussian statistics in the top figure. Similar to findings 
for turbulent convection in closed cylindrical vessels with solid walls the temperature field statistics 
deviates from Gaussian \cite{Emran2008}. The analysis can be refined. The inset of the top panel shows 
therefore vertical profiles of the plane- and time-averaged flatness, $F=\langle T^{\prime\, 4}\rangle_{A,t}/
\langle T^{\prime\,2}\rangle_{A,t}^2$. The flatness  differs clearly from the Gaussian value of 3 in all parts 
of the convection cell. In addition we plot the profile of the plane- and time-averaged skewness $S=\langle 
T^{\prime\,3}\rangle_{A,t}/\langle T^{\prime\,2}\rangle_{A,t}^{3/2}$ in the same inset. The magntiude of 
the skewness peaks at about $5\delta_T$ which is well inside the plume mixing zone \cite{Schumacher2008}. 
Both profiles agree also qualitatively with those by Kerr \cite{Kerr1996} and by Ref.
\cite{Emran2008} which have been conducted with no-slip top and bottom boundaries. In the bottom panel 
of Fig. \ref{fig4} we show the PDF of the temperature fluctuations in four different planes (see the legend).
The PDF in the midplane comes closest to a Gaussian profile. Our data suggest that the free-slip boundary 
conditions lead to smaller deviations from Gaussianity compared to the no-slip case. It should also be noted 
that for strong rotation of the cell about the vertical coordinate the temperature fluctuations are Gaussian for 
both, no-slip and free-slip boundary conditions, as reported by Julien {\it et al.} \cite{Julien1996}.  
  
\begin{figure}
\centerline{\includegraphics[scale=0.5]{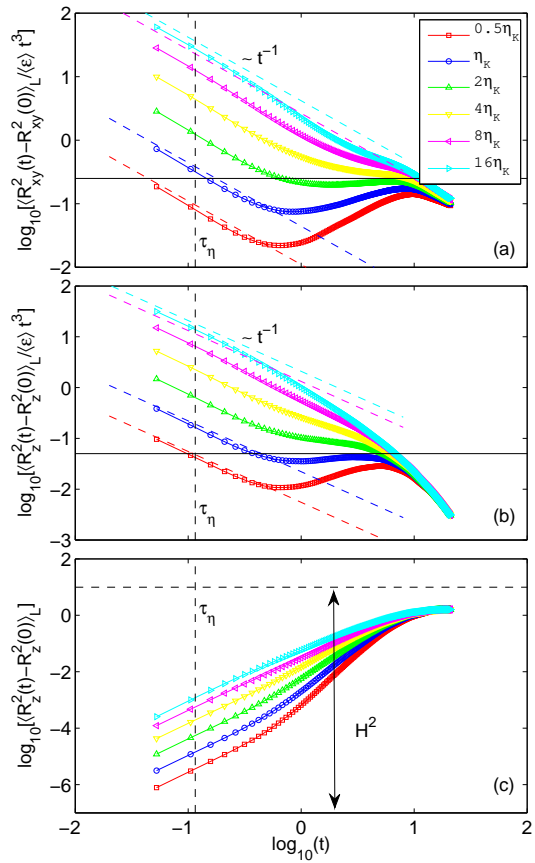}}
\caption{(Color online) Vertical and lateral particle pair dispersion as a function of the initial pair 
separation. (a) Compensated  lateral dispersion for different initial separations as indicated in the 
legend. The dashed lines correspond with Eq. (\ref{fit1}) for $R_0=\eta_K/2$ 
and $R_0=\eta_K$ and with Eq. (\ref{fit2}) for $R_0=8\eta_K$ and $R_0=16
\eta_K$.  (b) Compensated vertical dispersion. Again, Eqs. (\ref{fit1}) and (\ref{fit2}) are
fitted to the data for the 
same initial separations. (c) Same data as in (b) without compensation by $\langle\epsilon
\rangle t^3$. The vertical dispersion is constrained between the planes and levels thus off at 
larger times. The square of the cell height, $H^2$, and the Kolmogorov time scale, $\tau_{\eta}$, are indicated. 
All axes are given in decadic logarithm. Tracer pairs are seeded initially across the whole volume.}
\label{fig5}
\end{figure}

\section{Lagrangian particle dispersion}
\subsection{Two-particle dispersion}
The Eulerian framework analysis of turbulent convection demonstrated already that the flow is 
indeed inhomogeneous with respect to the vertical direction. Furthermore, we recall that the 
Lagrangian tracer motion is constrained between $z=0$ and $H$ since both walls cannot be 
penetrated. One motivation to study the dispersion in three-dimensional turbulent convection is 
therefore to verify if  the classical Richardson dispersion law \cite{Richardson1926} can be also 
observed for the present case. Recall that the Richardson dispersion law follows from a solution 
of a diffusion problem which assumes a homogeneous and isotropic turbulent state. It states that, 
given two particle tracks, ${\bm x}_2(t)$ and ${\bm x}_1(t)$ with  ${\bm x}_i=(x_i,y_i,z_i)$, the 
distance vector  ${\bm R}(t)={\bm x}_2(t)-{\bm x}_1(t)$  will follow 
\begin{equation}
\langle R^2(t)\rangle_L= g_{3d} \langle\epsilon\rangle t^3\,,
\end{equation}
where $g_{3d}$ is a universal constant of ${\cal O}(1)$. The symbol $\langle\cdot\rangle_L$ 
denotes an average over Lagrangian particle tracks. We decompose the relative tracer motion 
into a lateral and vertical contribution in order to separate homogeneous and inhomogeneous 
directions. The distance vector can be written as
\begin{equation}
{\bm R}(t)={\bm R}_{xy}(t)+R_z(t){\bm e}_z.
\end{equation}
The lateral two-particle dispersion is given by $\langle R^2_{xy}(t)-R^2_{xy}(0)\rangle_L$ 
where the average is taken over $6\times 10^5$ particle pairs. Here, 
\begin{equation}
{\bm R}_{xy}=[x_2(t)-x_1(t)]{\bm e}_x+[y_2(t)-y_1(t)]{\bm e}_y\,.
\end{equation}
Similarly, the vertical dispersion is given by  $\langle R^2_{z}(t)-R^2_{z}(0)\rangle_L$. 
The dispersion in each space direction would contribute with a weight of 1/3 in homogeneous 
isotropic turbulence.  In order to compare our pair dispersion results with the predictions for 
isotropic turbulence, we will introduce two weight factors, $C_{xy}=2/3$ for the lateral motion 
and $C_z=1/3$ for the vertical one.

Figure \ref{fig5} displays both dispersion processes with respect to time for six different 
initial pair separations as explained in section II C. The two-particle dispersion is given by a 
compensated plot in panels (a) and (b) of the figure. The graphs are normalized by $\langle
\epsilon\rangle t^3$ to capture a Richardson-like scaling as a plateau.  The initial ballistic 
behavior at small separations causes then an algebraic decay with $t^{-1}$.   Following 
Sawford {\it et al.} \cite{Sawford2008}, we fit the following two relations to our data at small times
\begin{equation}
\frac{\langle R^2_{m}(t)-R^2_{}(0)\rangle_L}{\langle\epsilon\rangle t^3}= 
\frac{C_{m}}{3}\left( \frac{R_{m}(0)}{\eta_K}\right)^2 \frac{\tau_{\eta}}{t} 
\label{fit1}
\end{equation}
if $R_{m}(0)\ll \eta_K$ and 
\begin{equation}
\frac{\langle R^2_{m}(t)-R^2_{m}(0)\rangle_L}{\langle\epsilon\rangle t^3}= 
\frac{11 C_{m}}{3} C \left(\frac{R_{m}(0)}{\eta_K}\right)^{2/3} \frac{\tau_{\eta}}{t}  
\label{fit2}
\end{equation}
if $\eta_K\ll R_{m}(0)\ll L$. Here $L$ is the outer scale of turbulence and 
$C\approx 2$ \cite{Sawford2008}. Index $m$ 
stands for the lateral terms, $xy$, or the vertical term, $z$. The agreement with (\ref{fit1}) for the initial 
Kolmogorov and sub-Kolmogorov separations is reasonable. For larger initial separations we use 
(\ref{fit2}). The larger the initial separation the better agree prediction and data.
We fitted the two smallest and  largest initial separations only. None of the initial 
separations is neither much  smaller nor much larger than the Kolmogorov scale which explains
the slight deviations of the numerical results from the laws (\ref{fit1}) and (\ref{fit2}). 
\begin{figure}
\centerline{\includegraphics[scale=0.5]{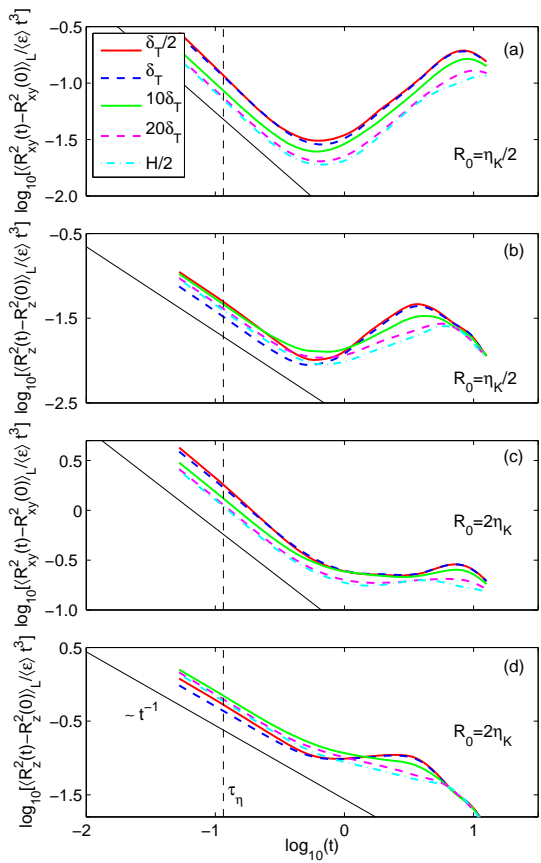}}
\caption{(Color online) Vertical and lateral particle pair dispersion as a function of the initial
vertical seeding position $z_0$. (a) Compensated  lateral dispersion for different initial heights $z_0$ 
as indicated in the legend (holds for all four figures). The initial tracer separation is $R_0=\eta_K/2$. (b) Compensated vertical 
dispersion.   The initial separation is also $R_0=\eta_K/2$. (c) Compensated  lateral dispersion.  
The initial separation is now $R_0=2\eta_K$. (d) Compensated vertical dispersion. Again 
$R_0=2\eta_K$. All axes are given in decadic logarithm and the Kolmogorov time scale is 
indicated by a dashed line.  The solid line follows $\sim t^{-1}$ in all plots.}
\label{fig6}
\end{figure}
 
As discussed for example in Refs. \cite{Bourgoin2006,Sawford2008}, the establishment of a 
Richardson-like regime depends sensitively on the initial separation between the tracers.  Indeed, 
for one of the six different separations the lateral dispersion curve passes through a small plateau 
with a Richardson constant $g_{xy}\approx0.25$. This is observed for an initial separation of 
$R_{xy}(0)=2\eta_K$, (see solid line in Fig. \ref{fig5}(a)). The re-translation of the proportionality constant
$g_{xy}$ into a three-dimensional homogeneous isotropic turbulence case is obtained by 
\begin{equation}
\tilde{g}_{xy}=\frac{g_{xy}}{C_{xy}}=\frac{3}{2} g_{xy}\approx 0.375\,.
\end{equation}
The proportionality constant is smaller than the value $g_{3d}\approx 0.5-0.6$ for homogeneous
isotropic turbulence  \cite{Boffetta2002,Biferale2005,Sawford2008}.  In Ref. \cite{Schumacher2008},  
it was already shown that the PDF of the lateral particle pair distance  can be fitted to the stretched 
exponential form of  Richardson \cite{Richardson1926}, however not to  the Gaussian shape as 
suggested by Batchelor \cite{Batchelor1950}.

Figures \ref{fig5}(b) and (c) display the vertical dispersion. In panel (b), we repeat the compensated 
plot of panel (a) and show the fits to (\ref{fit1}).  A plateau is observed now for initial separations between 
$1\eta_K$ and $2\eta_K$. The resulting constant is $g_z\approx 0.05$ (see solid line in Fig. \ref{fig5}(b)). 
If one combines the lateral and vertical dispersion, the Richardson constant for the turbulent convection 
follows  to 
\begin{equation}
g_{xy}+g_z\approx 0.3\,,
\end{equation}
which is less than our earlier estimate of $\tilde{g}_{xy}\approx 0.375$ and  $g_{3d}\approx 0.5-0.6$. 
A smaller Richardson constant corresponds with a stronger correlated pair motion. Such behavior can 
be attributed to the presence  of  rising and falling thermal plumes - a feature that is absent in isotropic 
turbulence.  Additionally, it is known that the plumes can cluster and form a large scale circulation 
\cite{Ahlers2009}.  Figure \ref{fig5}(c) demonstrates that the vertical dispersion is constrained by the top 
and bottom planes. The vertical contribution $\langle R_z^2(t)-R^2_z(0)\rangle_L$ to the pair dispersion 
levels off. Eventually the lateral dispersion contributes solely to the long-time behavior.

The specifics of the present inhomogeneous flow is that not only the initial pair separation, but also the initial seeding
position is important. This brings us to the second series of particle dispersion studies where tracer pairs 
with fixed distance in different horizontal planes of the slab are seeded (see section II C for details) and 
shorter simulations for about half the duration are rerun. Figure \ref{fig6} summarizes our findings.
We picked five intial seeding heights: two in the boundary layer, two in the plume mixing zone \cite{Schumacher2008} and
the center plane. While panels (a) and (b) are for $R_0=\eta_K/2$, panels (c) and (d) are for $R_0=2\eta_K$.
The latter is the separation that yielded a short Richardson-like range in Fig. \ref{fig5}(a).

Figures \ref{fig6}(a) and (c) show that the lateral dispersion curves of the tracer subgroups differ in magnitude.
The local slope is however nearly the same for all subsets. The situation is slightly different for the vertical dispersion: 
while the seeding in the center plane causes a gradual variation of the local slope of the dispersion curve
(see Figs. \ref{fig6}(b) and (d)),  the seeding in the thermal boundary layer leads to significant differences after the 
initial ballistic period. The same result is observed in Fig. \ref{fig6}(d). The reason is that the tracer pairs  
probe then the detachment of the boundary layer fragments to full extent. This is not the case when starting in the bulk
of the cell. It can also be observed that a plateau (which would imply Richardson-like scaling) depends sensitively 
on the intial separation and seeding height.            

To summarize this part,  Richardson-like dispersion appears for a very small range of scales in the present 
flow. Similar to previous studies,  we confirm that the initial pair dispersion depends sensitively on both, 
the initial pair separation and the initial vertical position of a tracer pair in the volume. The qualitative behavior 
of the tracer dispersion in the convection flow is very similar to that in homogeneous isotropic turbulence, 
given the same range of Reynolds numbers. The specifics of the turbulence, such as the particular driving 
mechanism or a present inhomogeneity, manifests however in the proportionality constant $g_{3d}$ and 
causes eventually quantitative deviations from homogeneous isotropic turbulence. Figure \ref{fig6}(c) illustrates this 
fact very nicely. The scatter of the plateaus can be interpreted as a measure of the sensitivity.    
    
\subsection{Multiparticle statistics}
The Lagrangian statistics of higher-order moments requires to follow more than two Lagrangian tracers 
simultaneously. In the following, we will focus to the four-particle case. The tracers are initially seeded 
at the edges of tetrahedra as discussed in section II C.  The distortion of such a small particle cluster 
by the turbulence has been studied for the pure hydrodynamic case in Refs. 
\cite{Chertkov1999,Pumir2000,Biferale2005,Cressman2004}. 
The original motivation for such analysis was to get a deeper geometrical insight into the formation of 
front-like structures in scalar turbulence: in the vicinity of steep scalar gradients small particle clusters 
become co-planar. Furthermore, since the cluster evolution probes the whole range of scales of turbulence, 
one hopes to disentangle systematically correlated large-scale advection from decorrelated small-scale 
motion. The presence of thermal plumes in convection will alter the deformation of the cluster at small 
times. It is however open, what will be observed in the long-time limit.
         
\begin{figure}
\centerline{\includegraphics[scale=0.5]{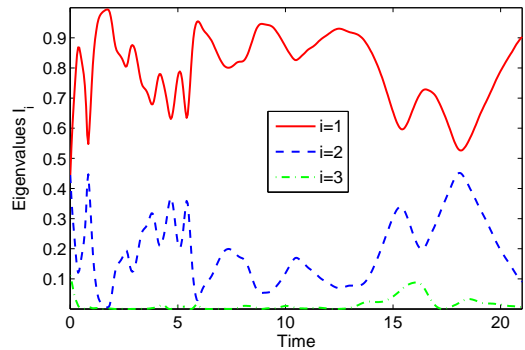}}
\caption{(Color online) Time evolution of the eigenvalues $I_i$ (see Eq. (\ref{eigenvalues}) for their 
definition) of one particular four-particle cluster. }
\label{fig7}
\end{figure}
The particle tracks ${\bm x}_1(t)$, ${\bm x}_2(t)$, ${\bm x}_3(t)$, and ${\bm x}_4(t)$ can be transformed 
into the center-of-mass coordinate
\begin{equation}  
{\bm r}(t) = \frac{1}{4} \sum_{i=1}^4 {\bm x}_i(t)\,,
\end{equation}
and the three relative coordinates (which are of interest here)
\begin{eqnarray}  
{\bm \rho}_1(t) &=& \frac{1}{\sqrt{2}} [{\bm x}_2(t)-{\bm x}_1(t)]\,\nonumber\\
{\bm \rho}_2(t) &=& \frac{1}{\sqrt{6}} [2 {\bm x}_3(t)-{\bm x}_1(t)-{\bm x}_2(t)]\,\nonumber\\
{\bm \rho}_3(t) &=& \frac{1}{\sqrt{12}} [3 {\bm x}_4(t)-{\bm x}_1(t)-{\bm x}_2(t)-{\bm x}_3(t)]\,.
\end{eqnarray}
The radius of gyration follows in this frame to $R_g=\sqrt{\rho_1^2+\rho_2^2+\rho_3^2}$. The shape evolution 
of the particle cluster is monitored by the following moment-of-inertia tensor
\begin{equation}  
g^{ab}=\sum_{i=1}^3 \rho_i^{a} \rho_i^{b}\,,
\end{equation}
where $a,b=x,y,z$ is the component index of the vector ${\bm \rho}_i$ and $i=1,2,3$. The real eigenvalues 
$g_1\ge g_2\ge g_3\ge 0$ quantify the shape of the particle cluster. Isotropic objects correspond
with $g_1=g_2=g_3$, cigar-shaped clusters with $g_1\gg g_2\approx g_3$ and pancake-shaped
clusters with $g_1\approx g_2\gg g_3$.  Figure \ref{fig7} shows the time evolution of the three eigenvalues
for one specific 4-particle cloud. The eigenvalues are normalized and given by  
\begin{equation}  
I_k=\frac{g_k}{\sum_{m=1}^3 g_m} \;\;\;\mbox{for}\;\;\;k=1, 2, 3\,. 
\label{eigenvalues}
\end{equation}
Thus $0\le I_k\le 1$. One can observe, that the eigenvalue variations become smoother with increasing time.
A convergence of the cluster to an almost coplanar object is observable for larger times, as quantified 
by the small value of $I_3$ . It will turn out now that this example displays a typical long-time behavior.

\begin{figure}
\centerline{\includegraphics[scale=0.5]{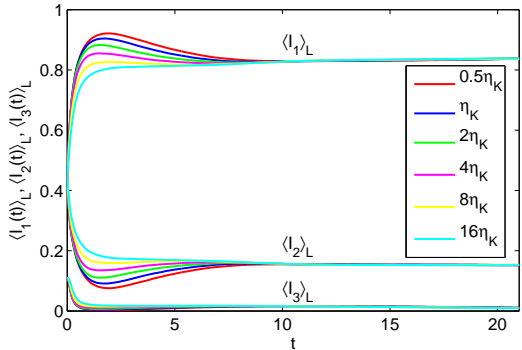}}
\caption{(Color online) Time evolution of the normalized eigenvalues of the moment-of-inertia tensor.
The average is taken over the whole ensemble of tetrahedra. For times $t>10 (=87\tau_{\eta})$ the data
converge to values $\langle I_3\rangle_L\to 0.84$, $\langle I_2\rangle_L\to 0.15$, and 
$\langle I_1\rangle_L\to 0.01$. The tetrahedra are seeded in the whole volume initially.}
\label{fig8}
\end{figure}
\begin{figure}
\centerline{\includegraphics[scale=0.5]{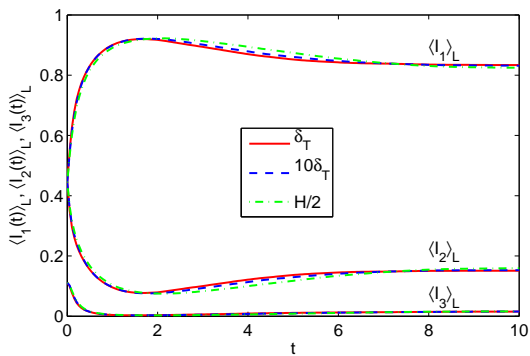}}
\caption{(Color online) Time evolution of the normalized eigenvalues of the moment-of-inertia tensor.
The average is taken again over the whole ensemble of $5\times 10^4$ tetrahedra for each case.
Four tracer particles form initially a tetrahedron with a sidelength of $\eta_K/2$. Three ensembles 
started in planes with $\delta$, 10$\delta_T$ and $H/2$ as given in the legend.}
\label{fig9}
\end{figure}
In Fig. \ref{fig8},  we show the time evolution of the Lagrangian ensemble average of the normalized 
eigenvalues. Data for different initial sidelengths of the tetrahedra are compared. The tetrahedra are seeded
across the whole volume.  Similar to the two-point measure, the initial deformation of the clusters depends 
sensitively on the sidelength of the tetrahedron. The smaller the initial sidelength the stronger
the initial stretching of the cluster to a cigar-shaped object (for $t\lesssim 15\tau_{\eta}$). After about $87 \tau_{\eta}$ 
all curves collapse and the mean values remain almost unchanged.
Our data yield $\langle I_1\rangle_L=0.84$,  $\langle I_2\rangle_L=0.15$, $\langle I_3\rangle_L=0.01$.
Suprisingly,  the obtained mean values are very close to the findings of Biferale {\it et al.} \cite{Biferale2005} 
and Hackl {\it et al.} \cite{Hackl2008} for homogeneous isotropic turbulence. In Ref. \cite{Biferale2005}, 
tetrahedra were excluded that had two points too close or too far of each other such that their values are not
directly comparable with the present ones. Recent three-dimensional particle tracking experiments by 
L\"uthi {\it et al.} \cite{Luethi2007} report an $\langle I_2\rangle_L$ which is also close to 0.16. 

Since the large scales are probed in the long-term limit by the particle clusters, 
this agreement suggests  that convective and isotropic turbulence on this scale do not differ significantly. 
The relative motion within a cluster is insensitive to whether the tracer particles are swept 
by large vortex structures or by large-scale circulation. Consequently, our findings suggest that 
the constrained vertical motion (and thus the inhomogeneity) is not important in the diffusive long-time limit
of the cluster dynamics.

One can expect that for times $t> 10^2\tau_{\eta}$, the tracers advance independently of each other.
The present long-time means are compared with the result of a joint Gaussian distribution of the relative 
coordinates, $p({\bm \rho_1}, {\bm \rho_2}, {\bm \rho_3}) \sim\exp[-(\rho_1^2+\rho_2^2+\rho_3^2)]$. This 
ansatz  results to $\langle I_1\rangle_G=0.75$, $\langle I_2\rangle_G=0.22$, and $\langle I_3\rangle_G=0.03$ 
for the three-dimensional case which is obtained by Monte-Carlo simulations \cite{Pumir2000}. The reported 
mean of $\langle I_2\rangle_L=0.15$ is smaller than the Gaussian value. 
 
Figure \ref{fig9} reports the dependence of the shape evolution from the initial position $z_0$. The effects 
remain small, but systematic. The closer the starting position of the tetrahedron to the boundary plane, the
faster it converges into the final quasistatic state. Again the stretching and deformation is most efficient when the
particle cluster passes through the mixing zone right above the tnermal boundary layer. 
     
\begin{figure}
\centerline{\includegraphics[scale=0.5]{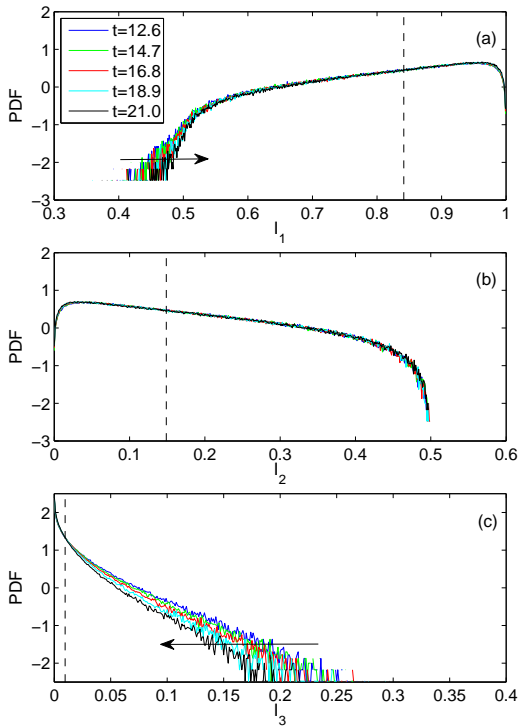}}
\caption{(Color online) Probability density functions of the normalized eigenvalues $I_k$ of the 
moment-of-inertia tensor (a) $I_1$, (b) $I_2$, and (c) $I_3$. The time instants at which the data 
have been analysed are for $t>87\tau_{\eta}$ as given in the legend in (a). The vertical dashed 
lines mark the long-time averages $\langle I_k\rangle_L$. The arrows in the upper and lower panel 
indicate that there is still a slight drift in the tails although the means in Fig. \ref{fig8} remain nearly 
unchanged. The tetrahedra are seeded again in the whole volume initially.}
\label{fig10}
\end{figure}

Figure \ref{fig10} displays the PDFs of $I_k$ for  five instants $t>87\tau_{\eta}$. The plots highlight two 
aspects.  First, there is still a big variety in the amplitudes of individual $I_k$ although their means 
remain almost unchanged. Second, a very slow drift in the tails is present which is indicated by the 
arrows in Fig. \ref{fig7}(a) and (c).

\section{Acceleration statistics} 
\begin{figure}
\centerline{\includegraphics[scale=0.5]{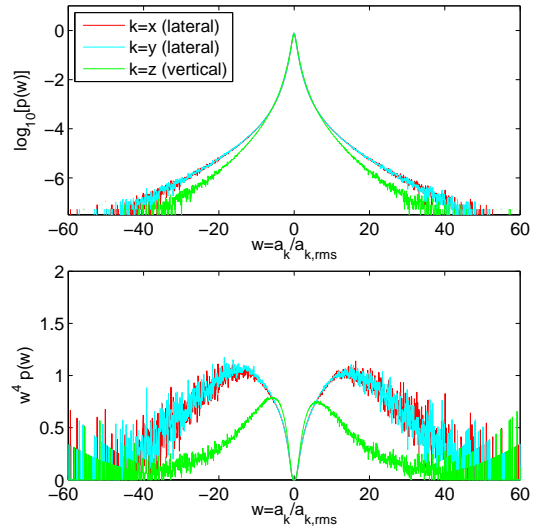}}
\caption{(Color online) Probability density function (PDF) of acceleration components $a_k$ with $k=x, y$ 
and $z$. Upper panel:  PDF plots. Each component is normalized by its corresponding root-mean-square 
value. Lower panel: Statistical convergence test for the 4th moment of the acceleration. The total number 
of events which as been included for the analysis are $4.4\times 10^8$.}
\label{fig11}
\end{figure}

The top panel of Fig. \ref{fig11} shows the PDFs of the three acceleration components. 
Each component is given in units of the corresponding root-mean-square value. As expected, 
the distributions of the two lateral components coincide. Table 1 provides the quantitative details 
of the acceleration statistics and lists for example the 
skewness $S(a_k)=\langle a_k^3\rangle/\langle a_k^2\rangle^{3/2}$ and the flatness 
$F(a_k)=\langle a_k^4\rangle/\langle a_k^2\rangle^{2}$   with $k=x, y$ or $z$. The numbers for
the lateral components are almost identical. The vertical acceleration component has a 
smaller flatness which is in line with a sparser tail of the corresponding PDF. The bottom panel
of the same figure provides the statistical convergence test of the fourth-order moments where the product
$w^4 p(w)$ is plotted vs. $w$ with $w=a_k/a_{k,rms}$.  It reflects the
fundamental difficulty to gather reliable statistics for higher-order moments in Lagrangian turbulence.
Recall that this analysis is conducted over a set of $4.4\times 10^8$ events. The area which is occupied 
by the scatter of the graphs in the tails of the PDFs determines the error bar of the fourth-order moment
(and consequently of the flatness). We checked that the second and third-order moments display 
almost no scatter (not shown). The issue of statistical convergence 
has been discussed for turbulence measurements in a swirling flow
\cite{Voth2002,Mordant2004} and in numerical simulations of homogeneous isotropic turbulence 
\cite{Toschi2005}. Our values for the flatness $F(a_k)$ of the lateral flatness are of the same magnitude 
as those reported in \cite{Voth2002}.  
\begin{figure}
\centerline{\includegraphics[scale=0.5]{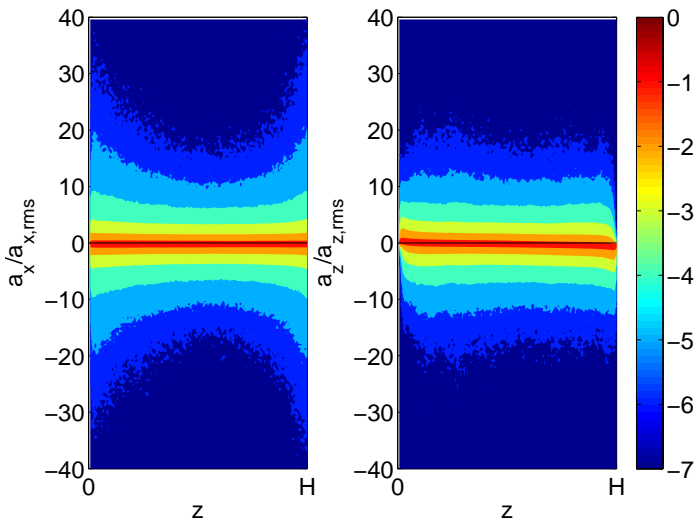}}
\caption{(Color online) Joint probability density function of acceleration components $a_i$ and the height $z$. 
Left:  component $a_x$. Right: component $a_z$. The contours are displayed units of the decadic logarithm.
The normalization of the acceleration component in both panels is as in Fig. \ref{fig11}.}
\label{fig12}
\end{figure}
\renewcommand{\arraystretch}{1.8}
\begin{table}
\begin{center}
\begin{tabular}{lccccc}
\hline\hline
           & $\sqrt{\langle a_k^2\rangle}$ & $\frac{\max(a_k)}{g}$ & $\frac{\min(a_k)}{g}$ 
           & $S(a_k)$ & $F(a_k)$\\
\hline
$a_x$ & 1.17 & 1001 &  -574 & -0.093    &  63.4 ($\pm$ 16)\\
$a_y$ & 1.16 & 559   &  -540 &  -0.156    &  64.4 ($\pm$ 16)\\
$a_z$ & 1.10 & 518   &  -680 &  -0.119    &  30.3 ($\pm$ 11)\\
\hline\hline
\end{tabular}
\caption{Root-mean-square values, total maximum/minimum amplitudes, skewness 
$S(a_k)=\langle a_k^3\rangle/\langle a_k^2\rangle^{3/2}$
and flatness $F(a_k)=\langle a_k^4\rangle/\langle a_k^2\rangle^2$ of the acceleration components. The error 
bars of $F(a_k)$ have been obtained by measuring the area of the scatter in the lower panel of Fig. \ref{fig11}. Maxima
and mininma are given in units of the gravity acceleration $g$.}
\end{center}
\label{table1}
\end{table}

Figure \ref{fig12} refines the statistical analysis of the acceleration components. Due to the vertical inhomogeneity, we
report the height dependence of the acceleration 
statistics for one lateral and the vertical component, respectively, and plot contours of the joint PDF 
$p(a_i, z)$.  The largest lateral accelerations and the fattest tails are found close to the top and bottom planes.  
It will turn out in the next section that the vorticity is concentrated in cyclones and anti-cyclones close to the
thermal boundary layer which can rationalize the large lateral accelerations. 
The support of the PDF decreases monotonically to the center plane. We will get back to this point later in the text
when discussing the role of the vertical vorticity component in connection with plume detachments. 
In contrast to the result for the lateral accelerations,  the support of the joint PDF $p(a_z, z)$ shows no 
significant variation with height. It shrinks to zero in the boundary planes (since is $u_z\equiv 0$) and grows rapidly 
up to about the thermal boundary layer thickness. The slight asymmetry of the inner contour lines (for the largest 
probability density levels) corresponds with the rising plumes which detach from the bottom plane at $z=0$ and have $a_z>0$ 
and with falling plumes at $z=H$ for which $a_z<0$. Note also that the support of all pdfs is the same in the center 
of the cell. This is consistent with the idea that the turbulence is close to isotropic far away from the isothermal walls.

The important result of this section is that there is differently strong intermittency for the vertical and lateral 
accelerations in thermal convection. It is caused by the higher level of intermittency in and close to the thermal boundary 
layer. As a consequence, we will have to take a closer look at the mechanisms of local heat transfer in the vicinity of the 
top and bottom planes. This is done in the next section.  
    
\begin{figure}
\centerline{\includegraphics[scale=0.4]{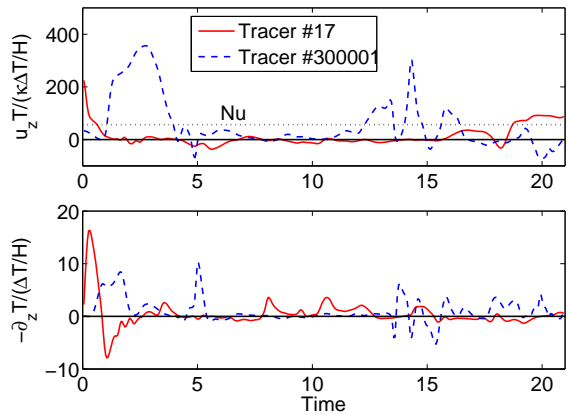}}
\caption{(Color online) Time traces of the convective (top) and conductive (bottom) heat transfer along two 
of the $1.2\times 10^6$ tracers. The Eulerian value of $Nu$ is indicated as a dotted line in the top panel. }
\label{fig13}
\end{figure}  

\section{Lagrangian convective and conductive heat flux}
\subsection{Lagrangian heat flux}
\begin{figure}
\centerline{\includegraphics[scale=0.6]{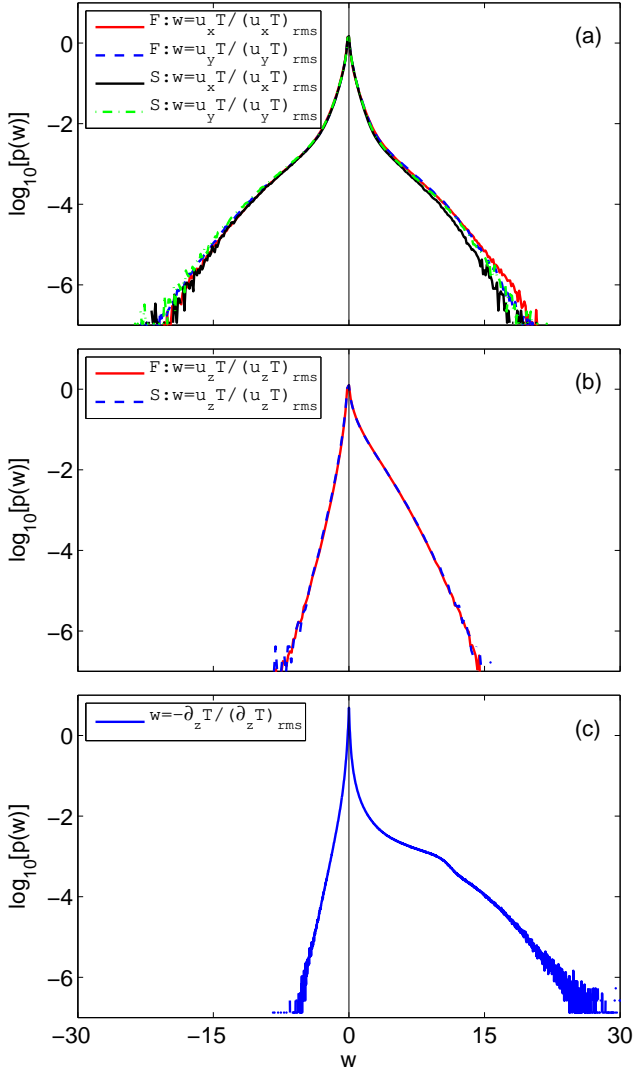}}
\caption{(Color online) Probability density functions (PDF) of the conductive and conductive heat flux contributions along the Lagrangian tracer tracks. (a) Lateral convective contributions $u_x T$ and $u_y T$. (b) Vertical convective contribution $u_z T$. (c) Conductive contribution $-\partial T/\partial z$. All quantities are normalized by their root-mean-square values. The analysis is conducted over two different data sets: the full record (F) and a third of it (S).}
\label{fig14}
\end{figure}

The local heat flux contributions can be probed in the Lagrangian frame of reference. 
We adopt therefore  definition (\ref{Nulocal}) and calculate the local Lagrangian conductive 
and convective flux contributions. They are given by
\begin{eqnarray}
Nu({\bm x})&=&Nu_{conv}({\bm x})+Nu_{cond}({\bm x})\nonumber\\
                    &=&\frac{H}{\kappa\Delta T} (u_z({\bm x})T({\bm x})-\kappa\partial_z T({\bm x}))\,,
\label{Nulocal1}                    
\end{eqnarray}
along the Lagrangian tracks ${\bm x}(t)$. Figure \ref{fig13} shows typical time traces of $Nu_{conv}$ 
and $Nu_{cond}$ along two tracers. They display a big variability with respect to time. Even negative 
values are possible for both contributions.  As expected, the convective term has a significantly larger 
magnitude than the conductive one. The latter is dominant in the thermal boundary layers where the 
thermal dissipation rate $\chi$ has the largest magnitude. Figure \ref{fig14} displays the PDFs of 
the convective and conductive contributions gathered along the tracks of the whole tracer ensemble. 
In addition, we display the results for the products $u_x T$ and $u_y T$. All quantities are shown in units 
of their root-mean-square values. While the PDFs for  $u_x T$ and $u_y T$ are symmetric, those of 
$u_z T$ and $\kappa\partial_z T$ are strongly skewed. This reflects the vertical net transfer of 
heat through the volume.  The Lagrangian average $Nu_L=\langle Nu({\bm x})\rangle_{L,t}$ results to 
\begin{equation}
Nu_L=1+\frac{H}{\kappa\Delta T} \langle u_z T\rangle_{L,t}\,.
\end{equation}
We have directly verified from the corresponding PDF in Fig. \ref{fig14}(c) that the mean of the 
Lagrangian conductive heat transfer is with 1.04 very close to one. Furthermore, it is found that 
$Nu_L<Nu$. This is in contrast  to the experimental findings for the smart 
particle probe of Gasteuil {\it et al.} \cite{Gasteuil2007} where $Nu<Nu_L$.  The reason for this  
difference might be due to the finite extension of the smart particle that exceeded the thickness 
of the thermal boundary layer. We have compared the result for the full record (F) with that of a
smaller subset (S) which is one third of set (F) and the results differed by 6.5 \% only (see also
Fig. \ref{fig14}). A time-resolved analysis shows that $Nu_L(t)=\langle Nu({\bm x},t)\rangle_{L}$
relaxes slowly to the Eulerian value of $Nu$. The tracers which have been seeded randomly or
at particular heights at the beginning have to pass a kind of ``thermalization" process.     

\begin{figure}
\centerline{\includegraphics[scale=0.45]{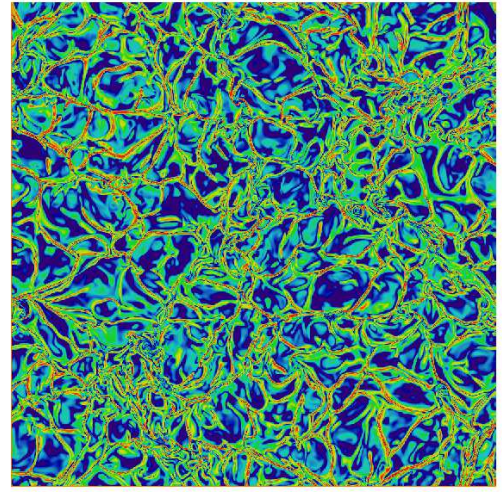}}
\caption{(Color online) Contour plot of the decadic logarithm of the thermal dissipation rate $\log_{10}
[\chi(x,y,z=\delta_T,t_0)]$. The logarithmic contour spacing is chosen  in order to 
highlight the small amplitude events.}
\label{fig15}
\end{figure}

\begin{figure}
\centerline{\includegraphics[scale=0.45]{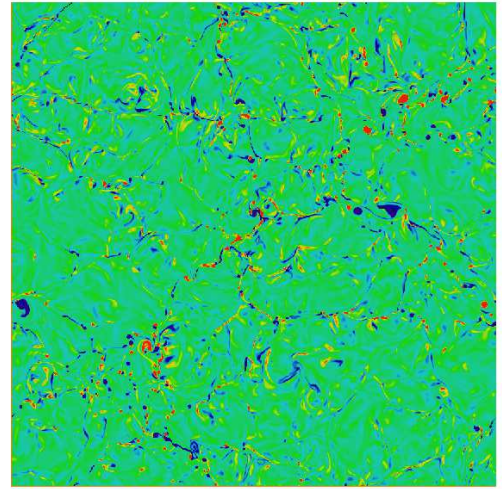}}
\caption{(Color online) Contour plot of the vertical vorticity component $\omega_z(x,y,z=\delta_T,t_0)$. 
Data correspond to those in Fig. \ref{fig12}.}
\label{fig16}
\end{figure}

\begin{figure}
\centerline{\includegraphics[scale=0.4]{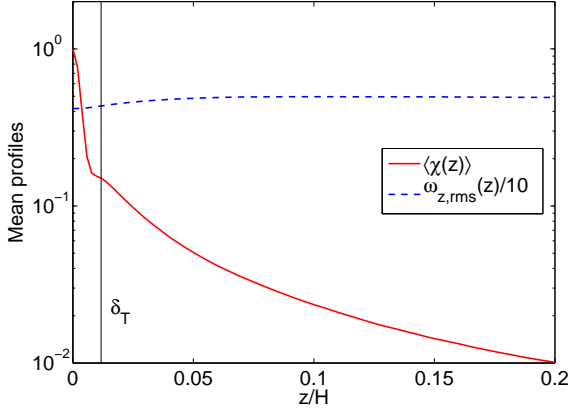}}
\caption{(Color online) Vertical profiles of the averaged thermal dissipation rate $\langle\chi(z)\rangle_{A,t}$
(solid line) and the root-mean-square of vertical vorticity component $\sqrt{\langle\omega^2(z)\rangle_{A,t}}$ 
(dashed line). The latter is divided by a factor of 10.} 
\label{fig17}
\end{figure}

Our result sheds interesting light on the joint velocity-temperature sampling properties of the Lagrangian tracers. First, 
it is known that for the present geometry a large-scale circulations are present \cite{Hartlep2005,Reeuwijk2008}. 
Tracers will preferentially follow the circulation motion in the convection layer. The large-scale
circulation can carry a fraction of the total heat transport only, as has been analyzed recently with a 
Proper Orthogonal Decomposition \cite{Bailon2009}. Second, we 
observe high-amplitude events of the vertical vorticity $\omega_z=\partial_x u_y-\partial_y u_x$ close to the 
maxima of the thermal dissipation rate $\chi$. This is shown in Figs. \ref{fig15} and \ref{fig16} where contour 
plots of slice snapshots of $\chi$ and $\omega_z$ at a height $z=\delta_T$ are compared. The local maxima 
in the dissipation rate plot (\ref{fig15}) reproduce the skeleton of plume sheets. In their vicinity, we observe  
cyclones and anti-cyclones that are generated in connection with the detachment of plume fragments. This 
observation is in line with results in \cite{Shishkina2008} for the non-rotating and with \cite{Julien1996} for 
the rotating case.  Exactly these cyclones and anti-cyclones cause the large lateral positive and negative 
accelerations as seen in the PDFs in Fig. \ref{fig11}. Fig. \ref{fig17} plots vertical profiles of means of both quantities.
While the root-mean-square of the vertical vorticity component varies weakly, the thermal dissipation rate 
is strongly peaked in and close to the boundary layer. It is also known from studies in homogeneous isotropic turbulence 
\cite{Toschi2009} that the Lagrangian tracers are not very frequently trapped in the core of such vortex 
structures. 
\begin{figure}
\centerline{\includegraphics[scale=0.6]{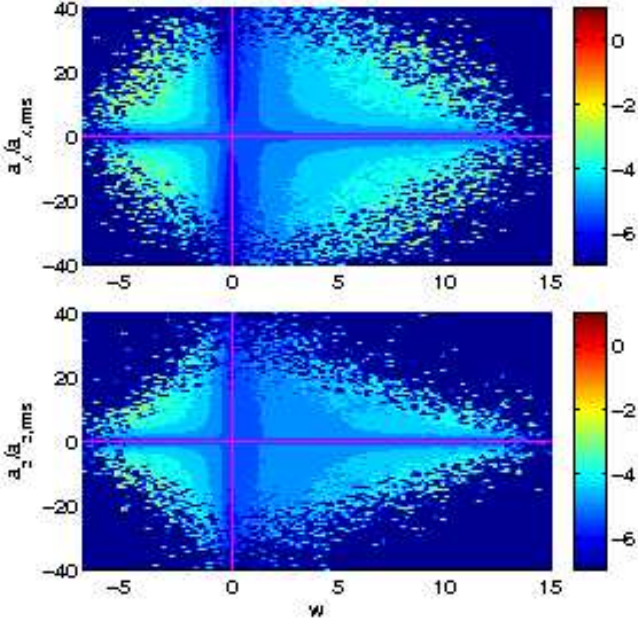}}
\caption{(Color online) Joint probability density function of the lateral acceleration (upper
panel) and vertical (lower panel) acceleration 
and heat flux contributions. Upper panel: $w=u_z T/(u_z T)_{rms}$. Lower panel: $w=u_z T/(u_z T)_{rms}$.
The contour spacing is in decadic logarithm.} 
\label{fig18}
\end{figure}

\subsection{Joint statistics of Lagrangian heat flux and acceleration}
The thermal plumes detach permantly from the thermal boundary layer and can be identified as 
regions in which the product  $u_z T^{\prime}>0$ \cite{Gasteuil2007,Schumacher2008}.  Here we
extend this analysis and study the correlations between the vertical velocity component and the (total) 
temperature in relation to the acceleration. Figure  \ref{fig18} displays the joint statistics 
of the vertical and lateral accelerations and the  products of velocity and temperature fluctuations, $u_z T$. In 
order to highlight the statistical correlation between the two variables of the joint PDF we divide the 
joint PDF by the two single quantity PDFs,
\begin{equation}
\Pi(a_i, u_z T)=\frac{p(a_i, u_zT)}{p(a_i) p(u_z T)}\,.
\end{equation}
The top panel of Fig. \ref{fig18} shows the joint statistics for $a_x$ and  $u_z T$. A pronunced maximum 
at larger accelerations and values of $u_z T$ is found. They can be related to coherent 
structures, such as vorticity tubes in the bulk of the slab or the cyclones/anti-cyclones in the boundary
layer. The correlation between vertical acceleration and the product 
$u_z T$ is  weaker. The local amplitudes of the joint PDF found at the outer boundaries of 
the support, at moderate acceleration amplitudes and larger values of $u_z T$. In Ref. 
\cite{Schumacher2008}, we reported the same behavior for  $p(a_z, u_z T^{\prime})$ and identified 
rising plumes $(a_z>0, u_z T^{\prime}>0)$ and  falling plumes  $(a_z<0, u_z T^{\prime}>0)$. Recirculations 
around rising and falling plumes have to form due to the incompressibility of the fluid. They were 
related to maxima in the halfplane $u_z T^{\prime}<0$. These are only some of the possible scenarios 
which can be assigned to strong correlations in the joint statistics. The firm conclusion which we can draw from the present 
analysis and the  one in \cite{Schumacher2008} is that plumes (and therefore the vertical convective 
flux events)  are {\em not} connected with the largest vertical accelerations. The detachment of plumes 
is a more gradual process.  

\begin{figure}
\centerline{\includegraphics[scale=0.6]{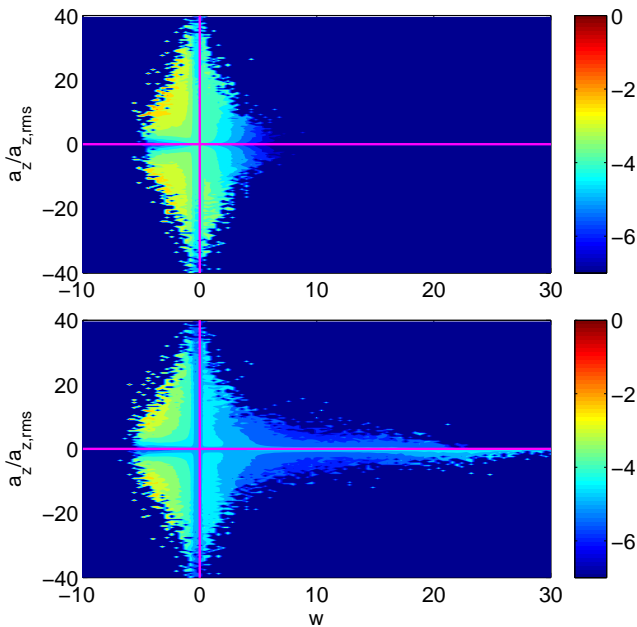}}
\caption{(Color online) Joint probability density function of the vertical acceleration component $a_z/a_{z,rms}$ 
and $w=-\partial_z T/(\partial_z T)_{rms}$. Top panel: Tracer positions in the bulk have been considered only. 
Bottom panel: Tracer positions in the whole volume are included. The contour spacing is in decadic logarithm.}
\label{fig19}
\end{figure}

We repeated the joint statistical analysis for the conductive part, $-\kappa\partial_z T({\bm x})$.  
Results are summarized in Fig. \ref{fig19}. The contour plots display now
\begin{equation}
\Pi(a_z, -\partial_z T)=\frac{p(a_z, -\partial_z T)}{p(a_z) p(-\partial_z T)}\,.
\end{equation}
We keep in mind from (\ref{Nulocal}) and (\ref{Nulocal1}) that upward conductive heat transport 
events have a negative sign, i.e. $-|\partial_z T|$. While the lower panel of Fig. \ref{fig19}  includes the Lagrangian data of 
the whole volume, the upper panel of the same figure excludes events in the thermal boundary layer 
up to the beginning of the plume mixing zones. This zone was identified and studied in \cite{Schumacher2008} 
and starts for the present parameter setting at a height H/16. The extended tail in the lower panel can thus be 
related to largest gradients (and thus largest thermal dissipation rate amplitudes) in and above the 
thermal boundary layer. The asymmetry to negative $a_z$ for the 
largest gradients can be interpreted as tracer  decelerations which are present when temperature gradients 
are formed. Negative accelerations (or decelerations) seem to be  frequently related to stagnation-point flow
topologies, those flows which can steepen the temperature field to large gradients.    

\section{Summary and discussion}
The focus of the present work was on Lagrangian aspects of turbulent convection. The results  can be 
summarized as follows. The study of pair and multiparticle dispersion yields qualitatively similar 
results compared to the homogeneous isotropic case. Although the scaling behavior of second order 
moments is sensitive to initial separations and seeding heights, initial separations close to the Kolmogorov 
length result in a short Richardson-like scaling range. Interestingly, we reproduce the same long-time 
limits for the particle cluster shapes as in isotropic turbulence despite turbulent convection is inhomogeneous. 
This limit deviates from the Gaussian value. Our results suggest that the dispersion 
laws can be obtained for more complex flows than isotropic homogeneous turbulence. The proportionality 
constants are however different which can  be attributed to qualitaively different turbulence structures
such as thermal plumes in convection. They affect the vertical dispersion more significantly than the lateral 
dispersion. 

The inhomogeneity of the convective turbulence manifests in less intermittent statistics of 
the vertical acceleration component compared to the lateral ones. Thermal plumes are not coupled with the
strongest accelerations. 

We find that the Lagrangian Nusselt number converges slowly to the Eulerian value.    A closer inspection 
of this point could be done in several steps: first, to disentangle the large-scale circulation from the turbulent 
background, as done in 
\cite{Bailon2009}, and to combine such a study with a Lagrangian analysis. Second, the present study indicates also that the 
Nusselt number relaxes faster with a growing number of tracers, in other words the sampling of joint velocity-temperature statistics improves. Our observation is also related to recent experimental and numerical studies at very large Rayleigh numbers  \cite{Amati2005,Niemela2008}  
in which the existence and growth of a so-called superconducting core is discussed, which is in line  with a decreasing
importance  of the large flow
circulation for growing Rayleigh numbers.  A decreasing importance of a coherent large-scale circulation 
might lead again to a faster convergence $Nu_L\rightarrow Nu$. More studies of the Lagrangian frame of high-Rayleigh 
number turbulence are thus necessary.    
  
\acknowledgements
The author wants  to thank Alain Pumir and Andr\'{e} Thess for comments and suggestions. 
This work is supported by the Heisenberg Program of the Deutsche Forschungsgemeinschaft (DFG) under
grant SCHU 1410/5-1. The author acknowledges support with computer time on the Blue Gene/P system JUGENE at the J\"ulich 
Supercomputing Centre J\"ulich (Germany) under grant HIL02. This work would not have been possible 
without the help by Mathias P\"utz (IBM Germany) to migrate the code onto the Blue Gene architecture 
and by Dmitry Pekurovsky (San Diego Supercomputer Center). The author thanks both of them.


\begin{thebibliography}{100}
\bibitem{Kadanoff2001} L. P. Kadanoff, Phys. Today {\bf 54}, 34 (2001).       

\bibitem{Ahlers2009} G. Ahlers, S. Grossmann, and D. Lohse, Rev. Mod. Phys., to be published.

\bibitem{Shishkina2007} O. Shishkina and C. Wagner, Phys. Fluids {\bf 19}, 085107 (2007).

\bibitem{Zhou2007} S.-Q. Zhou, C. Sun, and K.-Q. Xia, Phys. Rev. Lett. {\bf 98}, 074501 (2007).

\bibitem{Shishkina2008} O. Shishkina and C. Wagner, J. Fluid Mech. {\bf 599}, 383 (2008).

\bibitem{Emran2008} M. S. Emran and J. Schumacher, J. Fluid Mech. {\bf 611}, 13 (2008).

\bibitem{Kerr1996} R. M. Kerr, J. Fluid Mech. {\bf 310}, 139 (1996).

\bibitem{Niemela2000}  J. J. Niemela, L. Skrbek, K. R. Sreenivasan, and R. J. Donelly, 
Nature {\bf 404}, 837 (2000).

\bibitem{Amati2005} G.  Amati, K. Koal, F. Massaioli, K. R. Sreenivasan, and R. Verzicco,
Phys. Fluids {\bf 17}, 121701 (2005).

\bibitem{Funfschilling2005} D. Funfschilling, E. Brown, A. Nikolaenko, and G. Ahlers, 
J. Fluid Mech. {\bf 536}, 145 (2005).

\bibitem{Toschi2009} F. Toschi and E. Bodenschatz, Annu. Rev. Fluid Mech. {\bf 41}, 375 (2009). 

\bibitem{Schumacher2009} K. R. Sreenivasan and J. Schumacher, Phil. Trans. Roy. Soc., to be published.
 
\bibitem{Mann2000}  S. Ott and J. Mann, J. Fluid Mech. {\bf 422}, 207 (2000).

\bibitem{LaPorta2001} A. La Porta, G. A. Voth, A. M. Crawford, J. Alexander, and E. Bodenschatz,
                      Nature {\bf 409}, 1017 (2001).
                      
\bibitem{Guala2005} M. Guala, B. L\"uthi, A. Liberzon, A. Tsinober, and W. Kinzelbach, J. Fluid Mech. {\bf 533}, 339 (2005).

\bibitem{Mordant2001} N. Mordant, P. Metz, O. Michel und J.-F. Pinton, Phys. Rev. Lett. {\bf 87}, 214501 (2001).

\bibitem{Yeung2002} P. K. Yeung, Annu. Rev. Fluid Mech. {\bf 34}, 115 (2002). 

\bibitem{Boffetta2002} G. Boffetta and I. M. Sokolov, Phys. Rev. Lett. {\bf 88}, 094501 (2002).

\bibitem{Biferale2005} L. Biferale, G. Boffetta, A. Celani, B. J. Devinish, A. Lanotte, and F. Toschi, 
Phys. Fluids {\bf 17}, 115101 (2005). 

\bibitem{Sawford2008} B. L. Sawford, P. K. Yeung, and J. F. Hackl, Phys. Fluids {\bf 20}, 065111 (2008).

\bibitem{Gasteuil2007} Y. Gasteuil, W. L. Shew, M. Gibert, F. Chill\'{a}, B. Castaing, and J.-F. Pinton, 
Phys. Rev. Lett. {\bf 93}, 234302 (2007).

\bibitem{Calzavarini2005} E. Calzavarini, D. Lohse, F. Toschi, and R. Tripiccione, Phys. Fluids {\bf 17}, 055107 (2005).

\bibitem{Calzavarini2006} E. Calzavarini, C. R. Doering, J. D. Gibbon, D. Lohse, A. Tanabe, and F. Toschi, 
Phys. Rev. E {\bf 73}, 035301 (2006).
  
\bibitem{Schumacher2008}  J. Schumacher, Phys. Rev. Lett. {\bf 100}, 134502 (2008).

\bibitem{Castaing1989} B. Castaing, G. Gunarante, F. Heslot, L. P. Kadanoff, A. Libchaber, S. Thomae, X.-Z. Wu, 
S. Zaleski, and G. Zanetti, J. Fluid Mech. {\bf 204}, 1 (1989).

\bibitem{Xia2002} S.-Q. Zhou and K.-Q. Xia, Phys. Rev. Lett. {\bf 89}, 184502 (2002).

\bibitem{Chertkov1999} M. Chertkov, A. Pumir, and B. I. Shraiman, Phys. Fluids {\bf 11}, 2394 (1999).

\bibitem{Pumir2000} A. Pumir, B. I. Shraiman, and M. Chertkov, Phys. Rev. Lett. {\bf 85}, 5324 (2000).


\bibitem{Schumacher2007} J. Schumacher and M. P\"utz, {\em Turbulence in laterally extended systems}, in Proceedings of the International Conference 
ParCo 2007, Eds. C. Bischof, M. B\"ucker, P. Gibbon, G. Joubert, T. Lippert, B. Mohr, F. Peters, IOS
Press, Amsterdam, 585 (2007).

\bibitem{Pekurovsky2008} http://www.sdsc.edu/us/resources/p3dfft/index.php

\bibitem{Julien1996} K. Julien, S. Legg, J. Mc Williams, and J. Werne, J. Fluid Mech. {\bf 322}, 243 (1996).

\bibitem{Bourgoin2006} M. Bourgoin, N. T. Ouelette, H. Xu, J. Berg, and E. Bodenschatz, Science {\bf 311}, 835 (2006).

\bibitem{Richardson1926} L. F. Richardson, Proc. Roy. Soc. London Ser. A {\bf 110}, 709 (1926).

\bibitem{Batchelor1950} G. K. Batchelor, Q. J. R. Meteorol. Soc. {\bf 76}, 133 (1950).


\bibitem{Cressman2004} J. R. Cressman, W. I. Goldburg, and J. Schumacher, Europhys. Lett. {\bf 66}, 219 (2004).

\bibitem{Hackl2008} J. F. Hackl, P. K. Yeung, B. L. Sawford, and M. S. Borgas, Bull. Am. Phys. Soc. {\bf 53} (15), 298 (2008).

\bibitem{Luethi2007} B. L\"uthi, S. Ott, J. Berg, and J. Mann, J. Turb. {\bf 8}, 45 (2007). 

\bibitem{Voth2002}  G. A. Voth, A. La Porta, A. M. Crawford, J. Alexander, and E. Bodenschatz, J. Fluid Mech. {\bf 469}, 121 (2002).

\bibitem{Mordant2004}  N. Mordant, A. M. Crawford, and E. Bodenschatz, Physica D {\bf 193}, 245 (2004).  

\bibitem{Hartlep2005} T. Hartlep, A. Tilgner, and F. H. Busse, J. Fluid Mech. {\bf 544}, 309 (2005).

\bibitem{Reeuwijk2008} M. van Reeuwijk, H. J. J. Jonker, and K. Hanjali\'{c}, Phys. Rev. E {\bf 77}, 036311 (2008).

\bibitem{Bailon2009} J. Bailon-Cuba, M. S. Emran, and J. Schumacher, J. Fluid Mech., submitted (2009).

\bibitem{Toschi2005} F. Toschi, L. Biferale, G. Boffetta, A. Celani, B. J. Devenish, and A. Lanotte, J. Turb.
                                       {\bf 6}, 40 (2005).

\bibitem{Niemela2008} J. J. Niemela and K. R. Sreenivasan, Phys. Rev. Lett. {\bf 100}, 184502 (2008).

\end{thebibliography}
\end{document}